\documentclass[aps,pra,twocolumn,groupedaddress,superscriptaddress,bibnotes,amsfonts,
citeautoscript,a4paper]{revtex4-1}

\usepackage{graphicx}
\usepackage{dcolumn}
\usepackage{bm}
\usepackage{xcolor}
\usepackage{ulem}
\usepackage{hyperref}
\usepackage{amsmath}
\usepackage{cleveref}
\usepackage{float}
\usepackage{wasysym}

\usepackage{svg}
\usepackage{amssymb}
\usepackage{scalerel}
\usepackage{caption}
\usepackage{setspace}
 \usepackage{multirow}
 \usepackage{booktabs}
 \usepackage{siunitx}
 \usepackage{physics}
 \usepackage{amsmath}
\usepackage{orcidlink}

\hypersetup{colorlinks=true,allcolors=blue}

\begin{document}

\title{
Inducing room-temperature valley polarization of excitonic emission in transition metal dichalcogenide monolayers 
}

\author{Sergii~Morozov\,\orcidlink{0000-0002-5415-326X}}
\affiliation{
 POLIMA---Center for Polariton-driven Light--Matter Interactions, University of Southern Denmark, Campusvej 55, DK-5230 Odense M, Denmark
}
\email{semo@mci.sdu.dk}

\author{Torgom~Yezekyan\,\orcidlink{0000-0003-2019-2225}}
\affiliation{Center for Nano Optics, University of Southern Denmark, Campusvej 55, DK-5230~Odense~M, Denmark}

\author{Christian~Wolff\,\orcidlink{0000-0002-5759-6779}}
\affiliation{
 POLIMA---Center for Polariton-driven Light--Matter Interactions, University of Southern Denmark, Campusvej 55, DK-5230 Odense M, Denmark
}

\author{Sergey~I.~Bozhevolnyi\,\orcidlink{0000-0002-0393-4859}}
\affiliation{Center for Nano Optics, University of Southern Denmark, Campusvej 55, DK-5230~Odense~M, Denmark}
\affiliation{
 Danish Institute for Advanced Study, University of Southern Denmark, Campusvej 55, DK-5230 Odense M, Denmark
}

\author{N.~Asger~Mortensen\,\orcidlink{0000-0001-7936-6264}}
\affiliation{
 POLIMA---Center for Polariton-driven Light--Matter Interactions, University of Southern Denmark, Campusvej 55, DK-5230 Odense M, Denmark
}
\affiliation{
 Danish Institute for Advanced Study, University of Southern Denmark, Campusvej 55, DK-5230 Odense M, Denmark
}

\begin{abstract}
\vspace{0.35cm}
\textbf{Abstract.} 
The lowest energy states in transition metal dichalcogenide (TMD) monolayers follow valley selection rules, which have attracted vast interest due to the possibility of encoding and
processing of quantum information.
However, these quantum states are strongly affected by the  temperature-dependent intervalley scattering causing complete valley depolarization, which is hampering any practical applications of TMD monolayers at room temperature.
Therefore, for achieving clear and robust valley polarization in TMD monolayers one needs to suppress parasitic depolarization processes, which is the central challenge in the growing field of valleytronics.
Here, in electron-doping experiments on TMD monolayers, we demonstrate that strong doping levels beyond $10^{13}$~cm$^{-2}$ can induce 61\% and 37\% valley contrast at room temperature in tungsten diselenide and molybdenum diselenide monolayers, respectively.
Our results indicate that charged excitons in TMD monolayers can be utilized as quantum units in designing of practical valleytronic devices operating at 300~K.
\vspace{1cm}

\end{abstract}


\maketitle


\section{Introduction}

The ultimate thickness of transition metal dichalcogenide (TMD) monolayers and the convenient access to valley degrees of freedom are inspiring features for the next-generation electronic and optoelectronic devices.
Such valleytronic-based devices utilize the possibility to create imbalanced carrier populations in $K$ and $K'$ energy extrema (valleys) of the Brillouin zone by optical pumping, which is a convenient method for the encoding of quantum information~\cite{Mak2012,Zeng2012}.
These optically-initiated valley states can exhibit quantum coherence and thus are allowing for processing and manipulation of information, providing a material platform for quantum computing and quantum-information processing~\cite{Jones2013}.
However, valley depolarization processes may equilibrate the desired carrier population imbalance between $K$ and $K'$ valleys, practically leading to complete loss of encoded quantum information. 
Therefore, the control of parasitic depolarization processes is essential for unlocking the full power of valleytronics.

Valley polarization in TMD monolayers has been shown to be sensitive to temperature, magnetic field, mechanical strain, charge doping etc.~\cite{Zeng2012,Jones2013,Fen2019,Lundt2019,Zhen2023}. 
While cryogenic temperatures can suppress phonon-assisted valley depolarization processes~\cite{Zeng2012}, such conditions are impractical for real-life applications at ambient conditions.
Besides, cryogenic conditions are not a comprehensive remedy against other valley depolarization mechanisms as in case of the molybdenum diselenide monolayer (MoSe$_2$), which exhibits near-zero valley contrast even at 4~K~\cite{Wang2015}. 
To induce a high degree of valley polarization (DVP), the radiative relaxation processes with characteristic time $\tau_r$ should be faster than the valley depolarization time $\tau_v$, i.e., their ratio should be minimal in the phenomenological expression~\cite{Mak2012} 
\begin{equation}
    \textnormal{DVP} = \frac{P_0}{1+\tau_\textrm{r} / \tau_v},
    \label{eq1}
\end{equation}
where $P_0\leq 1$ is the initial laser-induced degree of polarization depending on the sample quality and environment.
High valley polarization at room temperature has been achieved through manipulation of $\tau_r$ (relative to $\tau_v$) by enhancing the emission rates with optical antennas~\cite{Yan2023,Lin2021}, or by introducing nonradiative recombination pathways in graphene-TMD heterostructures~\cite{Lorchat2018}.
A complementary approach relies on the screening effect of the electron-hole exchange interaction due to the carrier doping~\cite{Konabe2016,Miyauchi2018}, which in chalcogenide-alloyed monolayers allowed for a very high valley polarization of 50\% at room temperature~\cite{Liu2020}.
Similarly, direct electrostatic and optical doping techniques were demonstrated to enhance the valley polarization at low temperatures~\cite{Fen2019,Robert2021}, however such doping techniques do not provide sufficient doping levels to enable any practical valley contrast at room temperature. 

In this Letter, we demonstrate strong valley polarization of excitonic emission in electrochemically--doped TMD monolayers at room temperature. We achieve 61\% and 37\% valley contrast in tungsten diselenide (WSe$_2$) and molybdenum diselenide (MoSe$_2$) monolayers at high electron doping densities $>10^{13}$~cm$^{-2}$, respectively. 
We investigate the effect of electron doping on the characteristic emission and intervalley scattering times, which provides the control over the valley depolarization dynamics.
Our results pave the way for designing of practical valleytronic devices operating at room temperature. 

\begin{figure*}	
\includegraphics[width=0.97\linewidth]{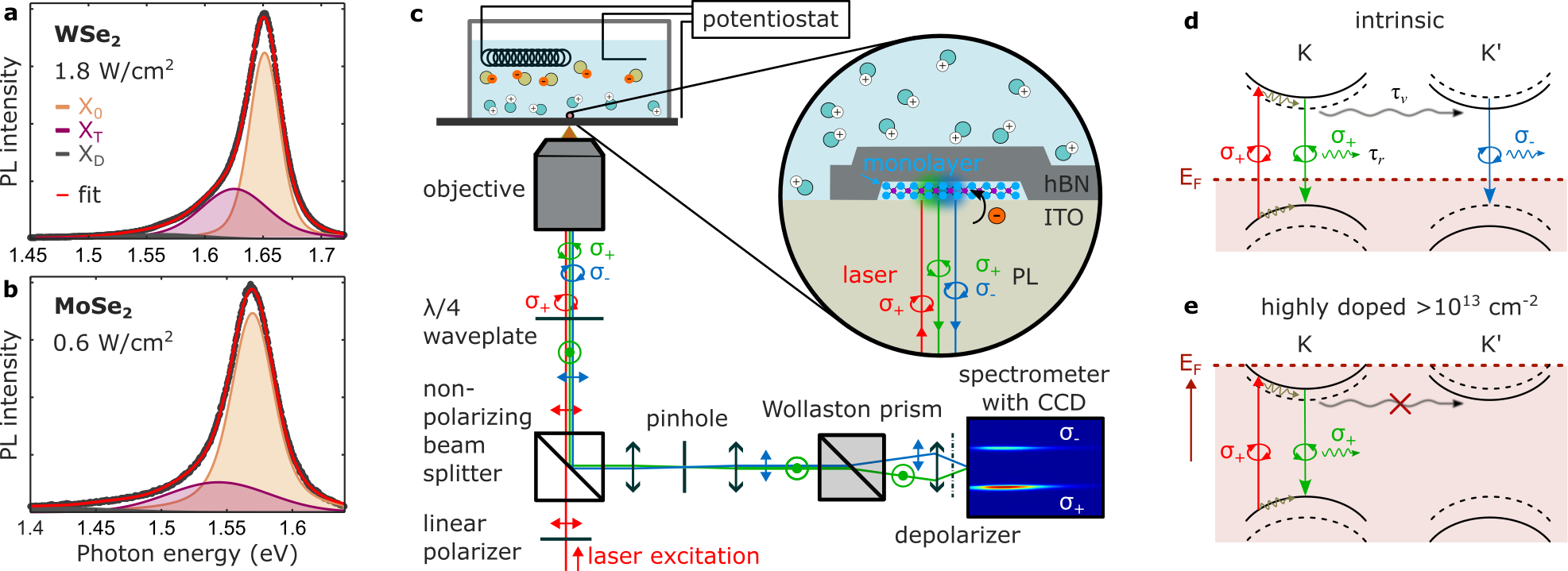}
\caption{\textbf{Charging excitons in electrochemical cell.}	
\textbf{a-b}~Photoluminescence spectra of MoSe$_2$ and WSe$_2$ monolayers at low laser power. 
The monolayers are sandwiched between an ITO substrate and a multilayer hBN flake. 
The orange and purple peaks represent the contribution of the neutral exciton X$_0$ and trion X$_T$, respectively, which are obtained by the spectral decomposition fit (red). 
\textbf{c}~Experimental setup for electron doping and polarization-sensitive detection of valley photoluminescence.
\textbf{d}~Schematic band structure of a TMD monolayer around $K$ and $K'$ points of the Brillouin zone. 
The red dashed line represents the Fermi level $E_F$, which for an intrinsic monolayer lies within the bandgap. 
The wavy grey arrow represents the intervalley scattering processes, the green and blue arrows -- radiative recombination generating right- ($\sigma_+$) or left-handed ($\sigma_-$) circularly polarized photons, respectively.
\textbf{e}~Application of a negative voltage bias in the electrochemical cell rises $E_F$ above the bottom of the conduction band causing strong electron doping, which leads to the suppression of intervalley scattering processes and cross-polarized photoluminescence.
}
\label{fig1}
\end{figure*}

 \section{Results and discussion}
\subsection{Samples and experimental setup}

We mechanically exfoliate WSe$_2$ and MoSe$_2$ bulk crystals into monolayers. 
All experiments presented were performed at room temperature of 293~K.
Representative photoluminescence spectra of exfoliated monolayers at low excitation power are shown in Fig.~\ref{fig1}a-b. 
Here, we use the spectral decomposition fitting with a three-peak Voigt function to distinguish contributions of the neutral exciton $X_0$ (orange) and negative trion $X_T$ (purple) emission~\cite{Morozov2021}. 
The third low-intensity peak (grey) is not of interest here, and, possibly, originates from the recombination through defect and impurity states.
We extract spectral parameters for $X_0$ and $X_T$ peak energies as well as their linewidth at half-intensity (Tab.~\ref{tab}).

\begin{table}[b]
\begin{tabular}{c|c|c|c|c}
         & $E_{X_0}$ (eV) & $\delta E_{X_0}$ (meV) & $E_{X_T}$ (eV) & $\delta E_{X_T}$ (meV) \\ \hline
WSe$_2$  & 1.670          & 35                     & 1.622          & 64 \\ 
\hline
MoSe$_2$ & 1.570          & 41                     & 1.543          & 118                          
\end{tabular}
\caption{\textbf{Photoluminescence of WSe$_2$ and MoSe$_2$ monolayers.} Results of spectral decomposition for peak energies of neutral exciton $E_{X_0}$ and trion $E_{X_T}$, and their linewidth at half-intensity $\delta E_{X_0,T}$.}
\label{tab}
\end{table}

We position the TMD monolayers in contact with an indium tin oxide (ITO) substrate (see the inset of Fig.~\ref{fig1}c), which serves as a working electrode of our custom-made electrochemical cell. 
Additionally, our monolayers are covered with hexagonal boron nitride (hBN) crystals for protecting them from the photo-chemical degradation due to the chemically-active electrolyte environment as well as direct Faradaic currents~\cite{Morozov2021}.
For the polarization measurements, the samples are tested with a right-handed  circularly polarized ($\sigma_+$) laser at off- and near-resonance energies (1.59~eV, 1.70~eV, and 1.81~eV).
Generated photoluminescence is analyzed with a polarization-resolving microscopy setup (Fig.~\ref{fig1}c), which is allowing for simultaneous detection of co- ($\sigma_+/\sigma_+$) and cross-polarized ($\sigma_+/\sigma_-$) emission on a charge-coupled device (CCD) camera of a spectrometer. 
The application of voltage bias in the electrochemical cell controls the position of the Fermi level $E_F$ in the TMD monolayer. 
At the neutral voltage bias of 0~V, $E_F$ is within the bandgap as in the case of an intrinsic monolayer (Fig.~\ref{fig1}d), while a negative bias can rise $E_F$ above the bottom of the conduction band, causing electron doping and filling of available states in the $K$ and $K'$ valleys of the Brillouin zone~(Fig.~\ref{fig1}e).  

\subsection{Optical doping}
First, we present the response of valley polarization to exciton charging using the optical doping technique (photocharging)~\cite{Fen2019}. 
High laser power (pump fluency) facilitates formation of charged excitons (trions, $X_T$), increasing the $X_T$ contribution in the photoluminescence signal. 
Here, we use right circularly polarized lasers at non-resonant energies of $E_{\rm exc}=1.81$~eV and 1.70~eV to excite the WSe$_2$ and MoSe$_2$ monolayers, respectively.

\begin{figure}[b!]	
\includegraphics[width=\linewidth]{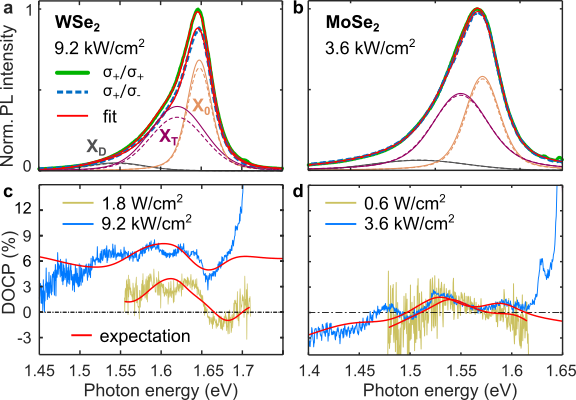}
\caption{\textbf{Valley-resolved photoluminescence in optically doped TMD monolayers.}	
\textbf{a-b}~At high power densities of $>1$~kW/cm$^{2}$, the WSe$_2$ monolayer exhibits slight valley polarization up to 9$\%$, while the MoSe$_2$ monolayer remains unpolarized.
The orange and purple peaks represent the neutral exciton X$_0$ and trion X$_T$ spectra extracted from the fits (red solid lines).
The solid and dashed lines correspond to the co- ($\sigma_+/\sigma_+$) and cross-polarized ($\sigma_+/\sigma_-$) detection channels. 
\textbf{c-d}~Degree of circular polarization  measured at low (yellow) and high (blue) excitation powers. 
The red lines are expected DOCP from the spectra fits in panels~\textbf{a-b}.
}
\label{fig2}
\end{figure}

Fig.~\ref{fig2}a-b presents polarization-resolved photoluminescence spectra of WSe$_2$ and MoSe$_2$ monolayers measured at high excitation power of $>1$~kW/cm$^{2}$, where the green and blue lines represent the co- and cross-polarized detection channels.  
We employ the spectra decomposition fit (red), extracting $X_0$ (orange) and $X_T$ (purple) peaks with respect to the polarization (solid and dashed lines for the co- and cross-polarized spectra).
We observe a slight valley polarization in the emission of the WSe$_2$ monolayer at high excitation power (Fig.~\ref{fig2}a,c), while the MoSe$_2$ monolayer exhibits no valley polarization regardless of excitation power (Fig.~\ref{fig2}b,d). 
We quantify this in Fig.~\ref{fig2}c-d (blue line) by the degree of circular polarization --- DOCP~$=(I_{\sigma_+}-I_{\sigma_-})/(I_{\sigma_+}+I_{\sigma_-})$ --- using the spectra from Fig.~\ref{fig2}a-b, respectively.
In comparison, the yellow line in Fig.~\ref{fig2}c-d represents a DOCP obtained at low excitation power. 
The increase of excitation power to 9.2~kW/cm$^2$ results in a growth of DOCP around the trion spectral line up to 9\%, which we attribute to the optical doping of WSe$_2$ monolayer.
We repeat the experiment on a different WSe$_2$ monolayer with a higher trion contribution to the signal, however obtaining similar low values of  valley polarization contrast. 
This demonstrates the limitations of the optical doping approach to achieve a high valley polarization at room temperature, while a further increase of excitation power may cause unwanted heating of the sample, parasitic defect and substrate emission, flourishing nonlinear processes, as well as irreversible photodegradation of a TMD monolayer.

We conclude here that the valley polarization in a WSe$_2$ monolayer (Fig.~\ref{fig2}a,c) mainly originates from the trion emission, which is in agreement with previous studies~\cite{Fen2019,Carmiggelt2020,Oliver2020}.
In contrast, an MoSe$_2$ monolayer exhibits no valley polarization in $X_0$ and $X_T$ spectral regions at low and high excitation powers, highlighting the material difference between the TMD monolayers. 
We also test our samples with a linearly polarized laser observing a near-zero degree of linear polarization (DOLP) in the $X_T$ spectral region, which is also independent of the excitation power. 
A linearly polarized excitation can generate quantum coherent valley states of the neutral exciton~\cite{Jones2013}, however we note that the detection of trion quantum coherence cannot be accessed in such an experiment and requires complementary experimental techniques~\cite{Hao2017}.

\begin{figure*}	
\includegraphics[width=\linewidth]{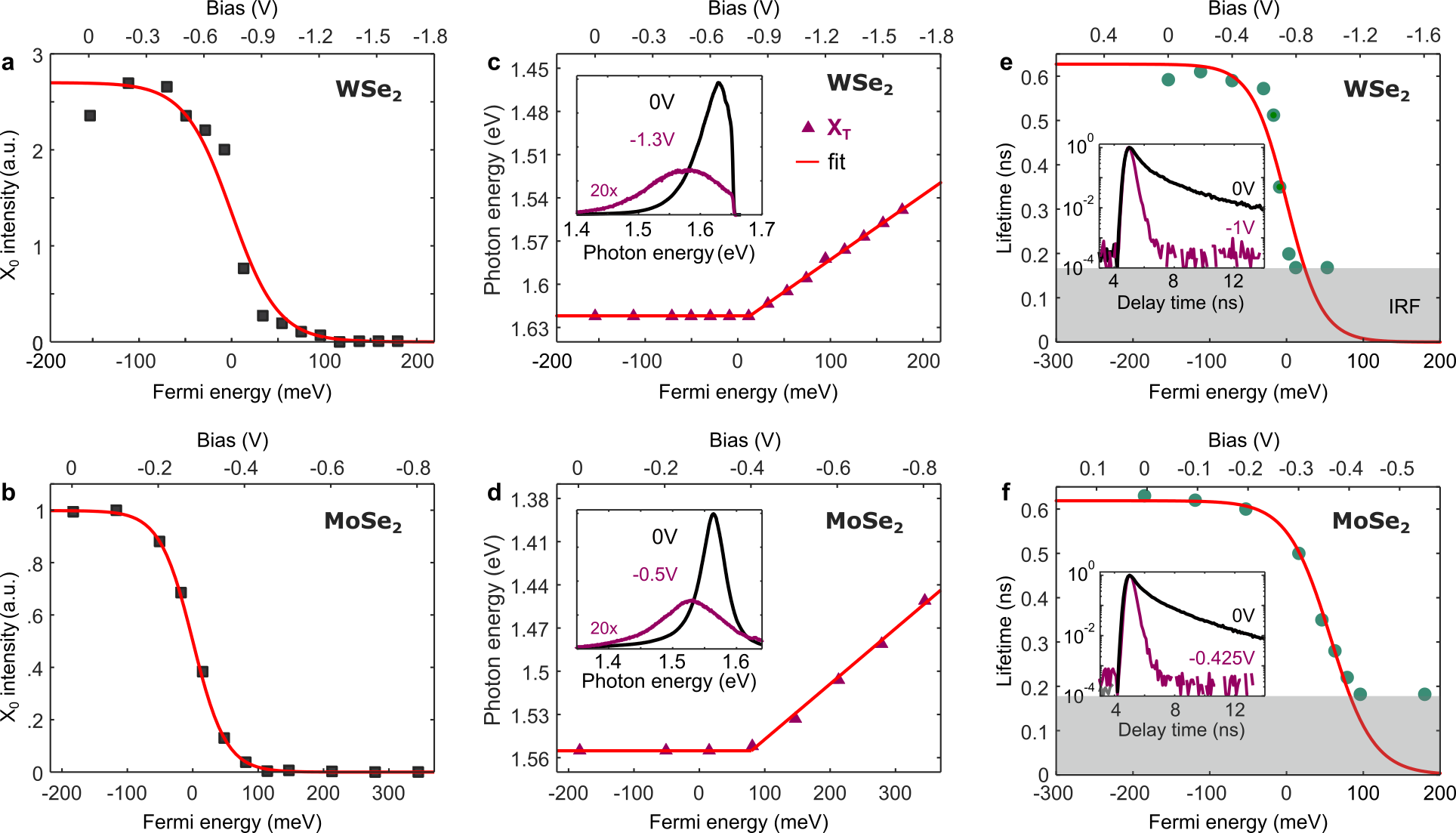}
\caption{\textbf{Photoluminescence in strongly doped TMD monolayers.}	
\textbf{a-b}~Intensity response of neutral exciton $X_0$ (black squares) to negative bias follows the Fermi--Dirac distribution (red).
\textbf{c-d}~Linear redshift of negative trion X$_T$ emission maximum (purple triangles) for Fermi levels above the bottom of conduction band. 
The insets show emission spectra at neutral (black) and negative (purple) bias, both monolayers are excited with a laser at 
$1.70$~eV.
\textbf{e-f}~Decrease of experimental lifetime at  negative bias (circles) follows the Fermi--Dirac distribution (red).
The insets show decay histograms at neutral (black) and negative (purple) bias.
The grey-shaded area represents the lifetime of experimental limit -- instrument response function (IRF).
}
\label{fig3}
\end{figure*}

\subsection{Electrochemical doping}

We now turn to discuss the manipulation of TMD monolayer photoluminescence via electrochemical doping. 
Evolution of photoluminescence intensities from WSe$_2$ and MoSe$_2$ monolayers during a linear scan of voltage bias (not shown) reveal that at negative bias the photoluminescence experiences a substantial drop in intensity. 
To characterize the change, we perform the spectral decomposition fitting, extracting the intensity evolution of the neutral exciton X$_0$ at negative bias (Fig.~\ref{fig3}a-b).
The X$_0$ intensity suppression dynamics follow the Fermi--Dirac distribution $f(E)=[\exp(-E/k_B T)+1]^{-1}$ at room temperature as $I(E_F)=I_0\, [1-f(E_F)]$ (red in Fig.~\ref{fig3}a-b), which allows us to translate the applied voltage bias to the Fermi energy with respect to the bottom of the conduction band, and  the electron doping density~\cite{Morozov2021}.
We attribute the drop in intensity to the filling of the monolayer conduction band with electrons, which causes depletion of absorption as well as suppression of neutral exciton emission. 

Fig.~\ref{fig3}c-d summarize the spectral response to negative bias. 
The insets of Fig.~\ref{fig3}c-d show photoluminescence spectra obtained at neutral and negative bias, demonstrating a substantial change of the emission energy. 
This is in line with our previous work~\cite{Morozov2021}, where we showed that the change in spectrum is due to the suppression of neutral exciton emission and linear redshift of the trion emission energy at increasing doping density. 
Besides, the trion emission energies of both WSe$_2$ and MoSe$_2$ start to redshift linearly once the Fermi level passes the bottom of the conduction band. 
We measure the maximal redshift in WSe$_2$ and MoSe$_2$ monolayers of 73~meV and 104~meV at -1.6~V and -0.8~V, respectively, which corresponds to electron doping density of $26\times 10^{12}$~cm$^{-2}$ and $48\times 10^{12}$~cm$^{-2}$. 

The suppression of $X_0$ intensity (Fig.~\ref{fig3}a-b) and the redshift in emission spectra (Fig.~\ref{fig3}c-d) at negative bias are accompanied by a reduction of experimental emission lifetime $\tau_r$ (Fig.~\ref{fig3}e-f). 
We note here that the reported experimental lifetimes do not represent the intrinsic exciton decay rate in WSe$_2$, being rather effective radiative lifetimes depending on various factors~\cite{Wangg2018}.
The insets of Fig.~\ref{fig3}e-f show decay histograms measured at neutral and negative bias, which we fit with a bi-exponential decay function to extract values of the experimental lifetime. 
Here, we observe a major reduction of $\tau_r$ (the quantitative estimation is beyond the resolution of our system of ca. 180~ps, grey-shaded region in Fig.~\ref{fig3}e-f), when the Fermi level is crossing the bottom of the conduction band.
Likewise the suppression of photoluminescence intensity in Fig.~\ref{fig3}a-b, we model the lifetime decrease with the Fermi--Dirac distribution (red in Fig.~\ref{fig3}e-f) as a function of the Fermi energy:
\begin{equation}
\tau_r=\tau_{r,0}\,[1-f(E_F)],
\label{eq2}
\end{equation}
where $\tau_{r,0}$ is the lifetime at 0~V bias.
We note here again the difference in the response of WSe$_2$ and MoSe$_2$ monolayers: while the redshift in spectra and the lifetime reduction happens for WSe$_2$ around zero Fermi energy in Fig.~\ref{fig3}c,e, the response of MoSe$_2$ is shifted to higher Fermi energies in Fig.~\ref{fig3}d,f. 

The reduction in experimental lifetime $\tau_r$ mainly stems from two processes: (i) formation of trions with faster recombination times $\tau_T$, and (ii) increasing probability of non-radiative Auger-related recombination characterized by $\tau_\textrm{nr}$~\cite{Carmiggelt2020}, that is $\tau_r^{-1}=\tau_T^{-1}+\tau_\textrm{nr}^{-1}$ in the spirit of Matthiessen's rule. 
We speculate that the strong electron doping enhances the radiative decay of excitonic emission, and if it is faster than the characteristic time of intervalley scattering processes at room temperature --- it can induce the
sought-after valley polarization contrast.

\begin{figure}	
\includegraphics[width=1\linewidth]{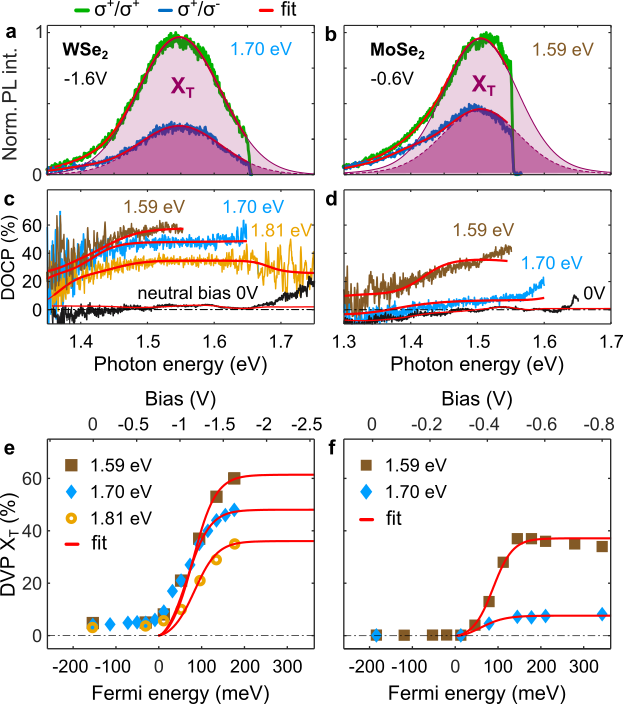}
\caption{\textbf{Inducing valley polarization in strongly doped TMD monolayers.}
\textbf{a-b} Polarization-resolved spectra acquired at negative bias and near-resonance excitation, 
where the green and blue lines correspond to the co- ($\sigma_+/\sigma_+$) and cross-polarized ($\sigma_+/\sigma_-$) detection channels, the purple-shaded peaks represent the trion contribution obtained from the spectral decomposition fit. 
\textbf{c-d} Corresponding DOCP at neutral (black) and negative (colored curves) bias, where colors correspond to the laser excitation energy. The red lines are expected DOCP from spectra fits as in panels \textbf{a-b}. 
\textbf{e-f} Evolution of the trion valley polarization in emission of WSe$_2$ and MoSe$_2$ monolayers at on- and off-resonant excitation. 
The transition to high valley polarization is fitted with Eq.~\ref{eqnM} (red).
}
\label{fig4}
\end{figure}

\subsection{Valley-polarization at strong electron doping}

Finally, we characterize valley polarization properties of WSe$_2$ and MoSe$_2$ monolayers at the negative bias voltages (laser power $<1$~kW/cm$^2$), which shift the Fermi level well above the bottom of conduction band. 
Photoluminescence of WSe$_2$ and MoSe$_2$ monolayers at neutral bias of 0~V are characterized by a low DOCP (black curves in Fig.~\ref{fig4}c-d), which are similar to what has been demonstrated in the optical doping experiments (Fig.~\ref{fig2}).
In contrast, Fig.~\ref{fig4}a-b present strongly polarized spectra of WSe$_2$ and MoSe$_2$ at high negative bias with dominating trion emission.
To quantify the trion valley polarization contrast, we employ the spectrum decomposition fit, extracting $X_T$ intensities in co- ($I_{\sigma_+}^T$) and cross-polarized ($I_{\sigma_-}^T$) spectra to calculate the experimental DVP of $X_T$ via $(I_{\sigma_+}^T-I_{\sigma_-}^T)/(I_{\sigma_+}^T+I_{\sigma_-}^T)$.
The purple-shaded areas in Fig.~\ref{fig4}a-b present the extracted trion intensity contributions $I_{\sigma_+}^T$ and  $I_{\sigma_-}^T$ to the spectra with respect to the polarization, which we used to calculate the experimental trion DVP of of 48\% and 36\% for WSe$_2$ and MoSe$_2$ monolayers, respectively.  
Furthermore, we test the polarization response to the excitation energy (Fig.~\ref{fig4}c-d), where the DOCP significantly increases for near-resonant excitation of the TMD monolayers.

\begin{figure}[b]	
\includegraphics[width=\linewidth]{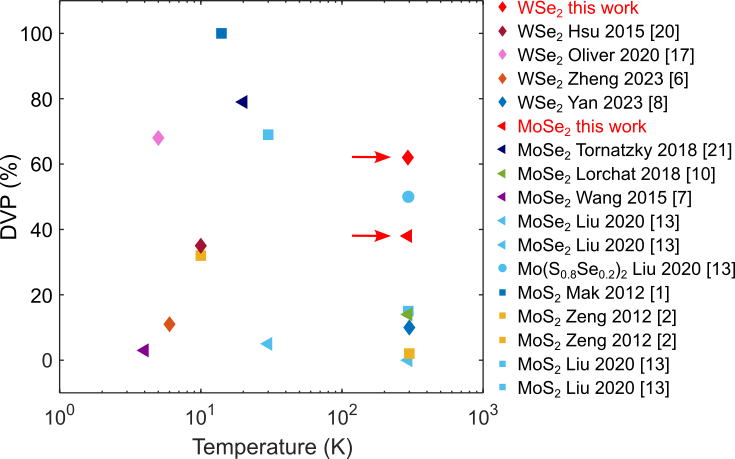}
\caption{\textbf{Valley polarization of TMD monolayers at cryogenic and room temperature.} Values obtained for WSe$_2$ and MoSe$_2$ monolayers in this work are highlighted by red color.	
Data points for WSe$_2$~($\blacklozenge$), MoSe$_2$~(\text{$\LHD$}), MoS$_2$~($\blacksquare$), and hybrid Mo(S$_{0.8}$Se$_{0.2}$$)_2$~($\CIRCLE$) are taken from references shown in the inset, where the same colors correspond to the same reference. 
}
\label{fig5}
\end{figure}

The dynamics of trion polarization at increasing negative bias is presented in Fig.~\ref{fig4}e-f.
Here, we measure spectra in a wide range of negative bias to extract the trion DVP as a function of bias (top scale) and Fermi energy (bottom scale).
Besides, the trion polarization dynamics follows a similar trend at off- and near-resonant laser excitation, experiencing a transition to strong valley polarization once the Fermi level is shifted above the bottom of conduction band ($E_F=0$).
We fit the transition to high valley polarization using Eq.~\ref{eq1} (red in Fig.~\ref{fig4}e-f), where we introduce a dependency on the Fermi energy for both characteristic lifetimes, i.e. 
\begin{equation}
    \textnormal{DVP}(E_F) = \frac{P_0}{1+\tau_r(E_F) / \tau_v(E_F)}.
    \label{eqnM}
\end{equation}
For the radiative lifetime $\tau_r$ dependency on $E_F$ we use Eq.~\ref{eq2}, while the valley depolarization time $\tau_v$ in the case of statically screened Coulomb potential with finite Thomas--Fermi wave vector can be expressed as~\cite{Konabe2016,Miyauchi2018,Liu2020}
\begin{equation}
\tau_v(E_F)=\tau_{v,0}\,[1-\text{exp}(-E_F/k_B T)]^2,
\end{equation}
where $\tau_{v,0}$ is the valley depolarization time at zero temperature, thus having the $\tau_{r,0}/\tau_{v,0}$ ratio as the only fitting parameter. 
The fits in Fig.~\ref{fig4}e-f (red) saturate in the strong-doping regime of $E_F>100$~meV ($>10^{13}$~cm$^{-2}$), and for the near-resonant excitation reach the trion DVP values of 61\% and 37\% for WSe$_2$ and MoSe$_2$ monolayers, respectively. 
We note here that the sole dependency of $\tau_v$ on $E_F$ in Eq.~\ref{eqnM} is not sufficient to unambiguously fit our results in Fig.~\ref{fig4}e-f, which highlights the importance of developing a more accurate model for the emission lifetime  $\tau_r(E_F)$ reduction accounting for many-body interactions at high electron concentrations. 

We compare our results with literature DVP  at cryogenic and room temperature conditions in Fig.~\ref{fig5}. 
Here, we plot DVP values for a range of TMD materials as well as their chalcogenide-alloyed hybrids.
With DVP values of 61\% for WSe$_2$ and 37\% for MoSe$_2$, we achieve the highest observed values to date at room temperature, which highlights the importance of the strong-doping regime for achieving a high DVP at ambient conditions.

\section{Conclusion}
In conclusion, we investigated valley polarization in WSe$_2$ and MoSe$_2$ monolayers under strong electron doping at room temperature. 
We achieved high trion DVP values of 61\% and 37\% in WSe$_2$ and MoSe$_2$ monolayers, respectively, under doping concentrations of over $10^{13}$~cm$^{-2}$.
The strong-doping regime allows for controlling the characteristic times of emission and intervalley scattering, which
prevents the intervalley scattering and induces the high valley polarization even at room temperature.
Our results demonstrate the importance of developing robust charge doping techniques to realize the strong-doping regime in TMD monolayers to enable the high valley polarization for valleytronic applications at room temperature.

\section{Acknowledgments}

We acknowledge stimulating discussions with Vladimir A.~Zenin. 
S.~M. acknowledges funding from the Marie Sk\l{}odowska-Curie Action (grant No.~101032967).
N.~A.~M. is a VILLUM Investigator supported by VILLUM FONDEN (grant No.~16498).
The Center for Polariton-driven Light-Matter Interactions (POLIMA) is funded by the Danish National Research Foundation (Project No.~DNRF165).

\bibliography{bibliography}

\begin{thebibliography}{19}%
\makeatletter
\providecommand \@ifxundefined [1]{%
 \@ifx{#1\undefined}
}%
\providecommand \@ifnum [1]{%
 \ifnum #1\expandafter \@firstoftwo
 \else \expandafter \@secondoftwo
 \fi
}%
\providecommand \@ifx [1]{%
 \ifx #1\expandafter \@firstoftwo
 \else \expandafter \@secondoftwo
 \fi
}%
\providecommand \natexlab [1]{#1}%
\providecommand \enquote  [1]{``#1''}%
\providecommand \bibnamefont  [1]{#1}%
\providecommand \bibfnamefont [1]{#1}%
\providecommand \citenamefont [1]{#1}%
\providecommand \href@noop [0]{\@secondoftwo}%
\providecommand \href [0]{\begingroup \@sanitize@url \@href}%
\providecommand \@href[1]{\@@startlink{#1}\@@href}%
\providecommand \@@href[1]{\endgroup#1\@@endlink}%
\providecommand \@sanitize@url [0]{\catcode `\\12\catcode `\$12\catcode
  `\&12\catcode `\#12\catcode `\^12\catcode `\_12\catcode `\%12\relax}%
\providecommand \@@startlink[1]{}%
\providecommand \@@endlink[0]{}%
\providecommand \url  [0]{\begingroup\@sanitize@url \@url }%
\providecommand \@url [1]{\endgroup\@href {#1}{\urlprefix }}%
\providecommand \urlprefix  [0]{URL }%
\providecommand \Eprint [0]{\href }%
\providecommand \doibase [0]{http://dx.doi.org/}%
\providecommand \selectlanguage [0]{\@gobble}%
\providecommand \bibinfo  [0]{\@secondoftwo}%
\providecommand \bibfield  [0]{\@secondoftwo}%
\providecommand \translation [1]{[#1]}%
\providecommand \BibitemOpen [0]{}%
\providecommand \bibitemStop [0]{}%
\providecommand \bibitemNoStop [0]{.\EOS\space}%
\providecommand \EOS [0]{\spacefactor3000\relax}%
\providecommand \BibitemShut  [1]{\csname bibitem#1\endcsname}%
\let\auto@bib@innerbib\@empty
\bibitem [{\citenamefont {Mak}\ \emph {et~al.}(2012)\citenamefont {Mak},
  \citenamefont {He}, \citenamefont {Shan},\ and\ \citenamefont
  {Heinz}}]{Mak2012}%
  \BibitemOpen
  \bibfield  {author} {\bibinfo {author} {\bibfnamefont {K.~F.}\ \bibnamefont
  {Mak}}, \bibinfo {author} {\bibfnamefont {K.}~\bibnamefont {He}}, \bibinfo
  {author} {\bibfnamefont {J.}~\bibnamefont {Shan}}, \ and\ \bibinfo {author}
  {\bibfnamefont {T.~F.}\ \bibnamefont {Heinz}},\ }\href {\doibase
  10.1038/nnano.2012.96} {\bibfield  {journal} {\bibinfo  {journal} {Nature
  Nanotechnology}\ }\textbf {\bibinfo {volume} {7}},\ \bibinfo {pages} {494}
  (\bibinfo {year} {2012})}\BibitemShut {NoStop}%
\bibitem [{\citenamefont {Zeng}\ \emph {et~al.}(2012)\citenamefont {Zeng},
  \citenamefont {Dai}, \citenamefont {Yao}, \citenamefont {Xiao},\ and\
  \citenamefont {Cui}}]{Zeng2012}%
  \BibitemOpen
  \bibfield  {author} {\bibinfo {author} {\bibfnamefont {H.}~\bibnamefont
  {Zeng}}, \bibinfo {author} {\bibfnamefont {J.}~\bibnamefont {Dai}}, \bibinfo
  {author} {\bibfnamefont {W.}~\bibnamefont {Yao}}, \bibinfo {author}
  {\bibfnamefont {D.}~\bibnamefont {Xiao}}, \ and\ \bibinfo {author}
  {\bibfnamefont {X.}~\bibnamefont {Cui}},\ }\href {\doibase
  10.1038/nnano.2012.95} {\bibfield  {journal} {\bibinfo  {journal} {Nature
  Nanotechnology}\ }\textbf {\bibinfo {volume} {7}},\ \bibinfo {pages} {490}
  (\bibinfo {year} {2012})}\BibitemShut {NoStop}%
\bibitem [{\citenamefont {Jones}\ \emph {et~al.}(2013)\citenamefont {Jones},
  \citenamefont {Yu}, \citenamefont {Ghimire}, \citenamefont {Wu},
  \citenamefont {Aivazian}, \citenamefont {Ross}, \citenamefont {Zhao},
  \citenamefont {Yan}, \citenamefont {Mandrus}, \citenamefont {Xiao},
  \citenamefont {Yao},\ and\ \citenamefont {Xu}}]{Jones2013}%
  \BibitemOpen
  \bibfield  {author} {\bibinfo {author} {\bibfnamefont {A.~M.}\ \bibnamefont
  {Jones}}, \bibinfo {author} {\bibfnamefont {H.}~\bibnamefont {Yu}}, \bibinfo
  {author} {\bibfnamefont {N.~J.}\ \bibnamefont {Ghimire}}, \bibinfo {author}
  {\bibfnamefont {S.}~\bibnamefont {Wu}}, \bibinfo {author} {\bibfnamefont
  {G.}~\bibnamefont {Aivazian}}, \bibinfo {author} {\bibfnamefont {J.~S.}\
  \bibnamefont {Ross}}, \bibinfo {author} {\bibfnamefont {B.}~\bibnamefont
  {Zhao}}, \bibinfo {author} {\bibfnamefont {J.}~\bibnamefont {Yan}}, \bibinfo
  {author} {\bibfnamefont {D.~G.}\ \bibnamefont {Mandrus}}, \bibinfo {author}
  {\bibfnamefont {D.}~\bibnamefont {Xiao}}, \bibinfo {author} {\bibfnamefont
  {W.}~\bibnamefont {Yao}}, \ and\ \bibinfo {author} {\bibfnamefont
  {X.}~\bibnamefont {Xu}},\ }\href {\doibase 10.1038/nnano.2013.151} {\bibfield
   {journal} {\bibinfo  {journal} {Nature Nanotechnology}\ }\textbf {\bibinfo
  {volume} {8}},\ \bibinfo {pages} {634} (\bibinfo {year} {2013})}\BibitemShut
  {NoStop}%
\bibitem [{\citenamefont {Feng}\ \emph {et~al.}(2019)\citenamefont {Feng},
  \citenamefont {Cong}, \citenamefont {Konabe}, \citenamefont {Zhang},
  \citenamefont {Shang}, \citenamefont {Chen}, \citenamefont {Zou},
  \citenamefont {Cao}, \citenamefont {Wu}, \citenamefont {Peimyoo},
  \citenamefont {Zhang},\ and\ \citenamefont {Yu}}]{Fen2019}%
  \BibitemOpen
  \bibfield  {author} {\bibinfo {author} {\bibfnamefont {S.}~\bibnamefont
  {Feng}}, \bibinfo {author} {\bibfnamefont {C.}~\bibnamefont {Cong}}, \bibinfo
  {author} {\bibfnamefont {S.}~\bibnamefont {Konabe}}, \bibinfo {author}
  {\bibfnamefont {J.}~\bibnamefont {Zhang}}, \bibinfo {author} {\bibfnamefont
  {J.}~\bibnamefont {Shang}}, \bibinfo {author} {\bibfnamefont
  {Y.}~\bibnamefont {Chen}}, \bibinfo {author} {\bibfnamefont {C.}~\bibnamefont
  {Zou}}, \bibinfo {author} {\bibfnamefont {B.}~\bibnamefont {Cao}}, \bibinfo
  {author} {\bibfnamefont {L.}~\bibnamefont {Wu}}, \bibinfo {author}
  {\bibfnamefont {N.}~\bibnamefont {Peimyoo}}, \bibinfo {author} {\bibfnamefont
  {B.}~\bibnamefont {Zhang}}, \ and\ \bibinfo {author} {\bibfnamefont
  {T.}~\bibnamefont {Yu}},\ }\href {\doibase 10.1002/smll.201805503} {\bibfield
   {journal} {\bibinfo  {journal} {Small}\ }\textbf {\bibinfo {volume} {15}},\
  \bibinfo {pages} {1805503} (\bibinfo {year} {2019})}\BibitemShut {NoStop}%
\bibitem [{\citenamefont {Lundt}\ \emph {et~al.}(2019)\citenamefont {Lundt},
  \citenamefont {Dusanowski}, \citenamefont {Sedov}, \citenamefont {Stepanov},
  \citenamefont {Glazov}, \citenamefont {Klembt}, \citenamefont {Klaas},
  \citenamefont {Beierlein}, \citenamefont {Qin}, \citenamefont {Tongay},
  \citenamefont {Richard}, \citenamefont {Kavokin}, \citenamefont
  {H\"{o}fling},\ and\ \citenamefont {Schneider}}]{Lundt2019}%
  \BibitemOpen
  \bibfield  {author} {\bibinfo {author} {\bibfnamefont {N.}~\bibnamefont
  {Lundt}}, \bibinfo {author} {\bibfnamefont {{\L}.}~\bibnamefont
  {Dusanowski}}, \bibinfo {author} {\bibfnamefont {E.}~\bibnamefont {Sedov}},
  \bibinfo {author} {\bibfnamefont {P.}~\bibnamefont {Stepanov}}, \bibinfo
  {author} {\bibfnamefont {M.~M.}\ \bibnamefont {Glazov}}, \bibinfo {author}
  {\bibfnamefont {S.}~\bibnamefont {Klembt}}, \bibinfo {author} {\bibfnamefont
  {M.}~\bibnamefont {Klaas}}, \bibinfo {author} {\bibfnamefont
  {J.}~\bibnamefont {Beierlein}}, \bibinfo {author} {\bibfnamefont
  {Y.}~\bibnamefont {Qin}}, \bibinfo {author} {\bibfnamefont {S.}~\bibnamefont
  {Tongay}}, \bibinfo {author} {\bibfnamefont {M.}~\bibnamefont {Richard}},
  \bibinfo {author} {\bibfnamefont {A.~V.}\ \bibnamefont {Kavokin}}, \bibinfo
  {author} {\bibfnamefont {S.}~\bibnamefont {H\"{o}fling}}, \ and\ \bibinfo
  {author} {\bibfnamefont {C.}~\bibnamefont {Schneider}},\ }\href {\doibase
  10.1038/s41565-019-0492-0} {\bibfield  {journal} {\bibinfo  {journal} {Nature
  Nanotechnology}\ }\textbf {\bibinfo {volume} {14}},\ \bibinfo {pages} {770}
  (\bibinfo {year} {2019})}\BibitemShut {NoStop}%
\bibitem [{\citenamefont {Zheng}\ \emph {et~al.}(2023)\citenamefont {Zheng},
  \citenamefont {Wu}, \citenamefont {Li}, \citenamefont {He}, \citenamefont
  {Liu}, \citenamefont {Wang}, \citenamefont {Wang}, \citenamefont {an~Duan},\
  and\ \citenamefont {Liu}}]{Zhen2023}%
  \BibitemOpen
  \bibfield  {author} {\bibinfo {author} {\bibfnamefont {H.}~\bibnamefont
  {Zheng}}, \bibinfo {author} {\bibfnamefont {B.}~\bibnamefont {Wu}}, \bibinfo
  {author} {\bibfnamefont {S.}~\bibnamefont {Li}}, \bibinfo {author}
  {\bibfnamefont {J.}~\bibnamefont {He}}, \bibinfo {author} {\bibfnamefont
  {Z.}~\bibnamefont {Liu}}, \bibinfo {author} {\bibfnamefont {C.-T.}\
  \bibnamefont {Wang}}, \bibinfo {author} {\bibfnamefont {J.-T.}\ \bibnamefont
  {Wang}}, \bibinfo {author} {\bibfnamefont {J.}~\bibnamefont {an~Duan}}, \
  and\ \bibinfo {author} {\bibfnamefont {Y.}~\bibnamefont {Liu}},\ }\href
  {\doibase 10.1364/ol.487201} {\bibfield  {journal} {\bibinfo  {journal}
  {Optics Letters}\ }\textbf {\bibinfo {volume} {48}},\ \bibinfo {pages} {2393}
  (\bibinfo {year} {2023})}\BibitemShut {NoStop}%
\bibitem [{\citenamefont {Wang}\ \emph {et~al.}(2015)\citenamefont {Wang},
  \citenamefont {Palleau}, \citenamefont {Amand}, \citenamefont {Tongay},
  \citenamefont {Marie},\ and\ \citenamefont {Urbaszek}}]{Wang2015}%
  \BibitemOpen
  \bibfield  {author} {\bibinfo {author} {\bibfnamefont {G.}~\bibnamefont
  {Wang}}, \bibinfo {author} {\bibfnamefont {E.}~\bibnamefont {Palleau}},
  \bibinfo {author} {\bibfnamefont {T.}~\bibnamefont {Amand}}, \bibinfo
  {author} {\bibfnamefont {S.}~\bibnamefont {Tongay}}, \bibinfo {author}
  {\bibfnamefont {X.}~\bibnamefont {Marie}}, \ and\ \bibinfo {author}
  {\bibfnamefont {B.}~\bibnamefont {Urbaszek}},\ }\href {\doibase
  10.1063/1.4916089} {\bibfield  {journal} {\bibinfo  {journal} {Applied
  Physics Letters}\ }\textbf {\bibinfo {volume} {106}},\ \bibinfo {pages}
  {112101} (\bibinfo {year} {2015})}\BibitemShut {NoStop}%
\bibitem [{\citenamefont {Yan}\ \emph {et~al.}(2023)\citenamefont {Yan},
  \citenamefont {Zhang}, \citenamefont {Li}, \citenamefont {Fang},
  \citenamefont {Jiang},\ and\ \citenamefont {Zhao}}]{Yan2023}%
  \BibitemOpen
  \bibfield  {author} {\bibinfo {author} {\bibfnamefont {Y.}~\bibnamefont
  {Yan}}, \bibinfo {author} {\bibfnamefont {X.}~\bibnamefont {Zhang}}, \bibinfo
  {author} {\bibfnamefont {X.}~\bibnamefont {Li}}, \bibinfo {author}
  {\bibfnamefont {H.}~\bibnamefont {Fang}}, \bibinfo {author} {\bibfnamefont
  {Y.}~\bibnamefont {Jiang}}, \ and\ \bibinfo {author} {\bibfnamefont
  {C.}~\bibnamefont {Zhao}},\ }\href {\doibase 10.1002/adfm.202213933}
  {\bibfield  {journal} {\bibinfo  {journal} {Advanced Functional Materials}\
  }\textbf {\bibinfo {volume} {33}},\ \bibinfo {pages} {2213933} (\bibinfo
  {year} {2023})}\BibitemShut {NoStop}%
\bibitem [{\citenamefont {Lin}\ \emph {et~al.}(2021)\citenamefont {Lin},
  \citenamefont {Wu}, \citenamefont {Akbari}, \citenamefont {Rossman},
  \citenamefont {Yeh},\ and\ \citenamefont {Atwater}}]{Lin2021}%
  \BibitemOpen
  \bibfield  {author} {\bibinfo {author} {\bibfnamefont {W.-H.}\ \bibnamefont
  {Lin}}, \bibinfo {author} {\bibfnamefont {P.~C.}\ \bibnamefont {Wu}},
  \bibinfo {author} {\bibfnamefont {H.}~\bibnamefont {Akbari}}, \bibinfo
  {author} {\bibfnamefont {G.~R.}\ \bibnamefont {Rossman}}, \bibinfo {author}
  {\bibfnamefont {N.-C.}\ \bibnamefont {Yeh}}, \ and\ \bibinfo {author}
  {\bibfnamefont {H.~A.}\ \bibnamefont {Atwater}},\ }\href {\doibase
  10.1002/adma.202104863} {\bibfield  {journal} {\bibinfo  {journal} {Advanced
  Materials}\ }\textbf {\bibinfo {volume} {34}},\ \bibinfo {pages} {2104863}
  (\bibinfo {year} {2021})}\BibitemShut {NoStop}%
\bibitem [{\citenamefont {Lorchat}\ \emph {et~al.}(2018)\citenamefont
  {Lorchat}, \citenamefont {Azzini}, \citenamefont {Chervy}, \citenamefont
  {Taniguchi}, \citenamefont {Watanabe}, \citenamefont {Ebbesen}, \citenamefont
  {Genet},\ and\ \citenamefont {Berciaud}}]{Lorchat2018}%
  \BibitemOpen
  \bibfield  {author} {\bibinfo {author} {\bibfnamefont {E.}~\bibnamefont
  {Lorchat}}, \bibinfo {author} {\bibfnamefont {S.}~\bibnamefont {Azzini}},
  \bibinfo {author} {\bibfnamefont {T.}~\bibnamefont {Chervy}}, \bibinfo
  {author} {\bibfnamefont {T.}~\bibnamefont {Taniguchi}}, \bibinfo {author}
  {\bibfnamefont {K.}~\bibnamefont {Watanabe}}, \bibinfo {author}
  {\bibfnamefont {T.~W.}\ \bibnamefont {Ebbesen}}, \bibinfo {author}
  {\bibfnamefont {C.}~\bibnamefont {Genet}}, \ and\ \bibinfo {author}
  {\bibfnamefont {S.}~\bibnamefont {Berciaud}},\ }\href {\doibase
  10.1021/acsphotonics.8b01306} {\bibfield  {journal} {\bibinfo  {journal}
  {{ACS} Photonics}\ }\textbf {\bibinfo {volume} {5}},\ \bibinfo {pages} {5047}
  (\bibinfo {year} {2018})}\BibitemShut {NoStop}%
\bibitem [{\citenamefont {Konabe}(2016)}]{Konabe2016}%
  \BibitemOpen
  \bibfield  {author} {\bibinfo {author} {\bibfnamefont {S.}~\bibnamefont
  {Konabe}},\ }\href {\doibase 10.1063/1.4961110} {\bibfield  {journal}
  {\bibinfo  {journal} {Applied Physics Letters}\ }\textbf {\bibinfo {volume}
  {109}},\ \bibinfo {pages} {073104} (\bibinfo {year} {2016})}\BibitemShut
  {NoStop}%
\bibitem [{\citenamefont {Miyauchi}\ \emph {et~al.}(2018)\citenamefont
  {Miyauchi}, \citenamefont {Konabe}, \citenamefont {Wang}, \citenamefont
  {Zhang}, \citenamefont {Hwang}, \citenamefont {Hasegawa}, \citenamefont
  {Zhou}, \citenamefont {Mouri}, \citenamefont {Toh}, \citenamefont {Eda},\
  and\ \citenamefont {Matsuda}}]{Miyauchi2018}%
  \BibitemOpen
  \bibfield  {author} {\bibinfo {author} {\bibfnamefont {Y.}~\bibnamefont
  {Miyauchi}}, \bibinfo {author} {\bibfnamefont {S.}~\bibnamefont {Konabe}},
  \bibinfo {author} {\bibfnamefont {F.}~\bibnamefont {Wang}}, \bibinfo {author}
  {\bibfnamefont {W.}~\bibnamefont {Zhang}}, \bibinfo {author} {\bibfnamefont
  {A.}~\bibnamefont {Hwang}}, \bibinfo {author} {\bibfnamefont
  {Y.}~\bibnamefont {Hasegawa}}, \bibinfo {author} {\bibfnamefont
  {L.}~\bibnamefont {Zhou}}, \bibinfo {author} {\bibfnamefont {S.}~\bibnamefont
  {Mouri}}, \bibinfo {author} {\bibfnamefont {M.}~\bibnamefont {Toh}}, \bibinfo
  {author} {\bibfnamefont {G.}~\bibnamefont {Eda}}, \ and\ \bibinfo {author}
  {\bibfnamefont {K.}~\bibnamefont {Matsuda}},\ }\href {\doibase
  10.1038/s41467-018-04988-x} {\bibfield  {journal} {\bibinfo  {journal}
  {Nature Communications}\ }\textbf {\bibinfo {volume} {9}},\ \bibinfo {pages}
  {2598} (\bibinfo {year} {2018})}\BibitemShut {NoStop}%
\bibitem [{\citenamefont {Liu}\ \emph {et~al.}(2020)\citenamefont {Liu},
  \citenamefont {del {\'{A}}guila}, \citenamefont {Liu}, \citenamefont {Zhu},
  \citenamefont {Han}, \citenamefont {Chaturvedi}, \citenamefont {Gong},
  \citenamefont {Yu}, \citenamefont {Zhang}, \citenamefont {Yao},\ and\
  \citenamefont {Xiong}}]{Liu2020}%
  \BibitemOpen
  \bibfield  {author} {\bibinfo {author} {\bibfnamefont {S.}~\bibnamefont
  {Liu}}, \bibinfo {author} {\bibfnamefont {A.~G.}\ \bibnamefont {del
  {\'{A}}guila}}, \bibinfo {author} {\bibfnamefont {X.}~\bibnamefont {Liu}},
  \bibinfo {author} {\bibfnamefont {Y.}~\bibnamefont {Zhu}}, \bibinfo {author}
  {\bibfnamefont {Y.}~\bibnamefont {Han}}, \bibinfo {author} {\bibfnamefont
  {A.}~\bibnamefont {Chaturvedi}}, \bibinfo {author} {\bibfnamefont
  {P.}~\bibnamefont {Gong}}, \bibinfo {author} {\bibfnamefont {H.}~\bibnamefont
  {Yu}}, \bibinfo {author} {\bibfnamefont {H.}~\bibnamefont {Zhang}}, \bibinfo
  {author} {\bibfnamefont {W.}~\bibnamefont {Yao}}, \ and\ \bibinfo {author}
  {\bibfnamefont {Q.}~\bibnamefont {Xiong}},\ }\href {\doibase
  10.1021/acsnano.0c02703} {\bibfield  {journal} {\bibinfo  {journal} {{ACS}
  Nano}\ }\textbf {\bibinfo {volume} {14}},\ \bibinfo {pages} {9873} (\bibinfo
  {year} {2020})}\BibitemShut {NoStop}%
\bibitem [{\citenamefont {Robert}\ \emph {et~al.}(2021)\citenamefont {Robert},
  \citenamefont {Park}, \citenamefont {Cadiz}, \citenamefont {Lombez},
  \citenamefont {Ren}, \citenamefont {Tornatzky}, \citenamefont {Rowe},
  \citenamefont {Paget}, \citenamefont {Sirotti}, \citenamefont {Yang},
  \citenamefont {Tuan}, \citenamefont {Taniguchi}, \citenamefont {Urbaszek},
  \citenamefont {Watanabe}, \citenamefont {Amand}, \citenamefont {Dery},\ and\
  \citenamefont {Marie}}]{Robert2021}%
  \BibitemOpen
  \bibfield  {author} {\bibinfo {author} {\bibfnamefont {C.}~\bibnamefont
  {Robert}}, \bibinfo {author} {\bibfnamefont {S.}~\bibnamefont {Park}},
  \bibinfo {author} {\bibfnamefont {F.}~\bibnamefont {Cadiz}}, \bibinfo
  {author} {\bibfnamefont {L.}~\bibnamefont {Lombez}}, \bibinfo {author}
  {\bibfnamefont {L.}~\bibnamefont {Ren}}, \bibinfo {author} {\bibfnamefont
  {H.}~\bibnamefont {Tornatzky}}, \bibinfo {author} {\bibfnamefont
  {A.}~\bibnamefont {Rowe}}, \bibinfo {author} {\bibfnamefont {D.}~\bibnamefont
  {Paget}}, \bibinfo {author} {\bibfnamefont {F.}~\bibnamefont {Sirotti}},
  \bibinfo {author} {\bibfnamefont {M.}~\bibnamefont {Yang}}, \bibinfo {author}
  {\bibfnamefont {D.~V.}\ \bibnamefont {Tuan}}, \bibinfo {author}
  {\bibfnamefont {T.}~\bibnamefont {Taniguchi}}, \bibinfo {author}
  {\bibfnamefont {B.}~\bibnamefont {Urbaszek}}, \bibinfo {author}
  {\bibfnamefont {K.}~\bibnamefont {Watanabe}}, \bibinfo {author}
  {\bibfnamefont {T.}~\bibnamefont {Amand}}, \bibinfo {author} {\bibfnamefont
  {H.}~\bibnamefont {Dery}}, \ and\ \bibinfo {author} {\bibfnamefont
  {X.}~\bibnamefont {Marie}},\ }\href {\doibase 10.1038/s41467-021-25747-5}
  {\bibfield  {journal} {\bibinfo  {journal} {Nature Communications}\ }\textbf
  {\bibinfo {volume} {12}},\ \bibinfo {pages} {5455} (\bibinfo {year}
  {2021})}\BibitemShut {NoStop}%
\bibitem [{\citenamefont {Morozov}\ \emph {et~al.}(2021)\citenamefont
  {Morozov}, \citenamefont {Wolff},\ and\ \citenamefont
  {Mortensen}}]{Morozov2021}%
  \BibitemOpen
  \bibfield  {author} {\bibinfo {author} {\bibfnamefont {S.}~\bibnamefont
  {Morozov}}, \bibinfo {author} {\bibfnamefont {C.}~\bibnamefont {Wolff}}, \
  and\ \bibinfo {author} {\bibfnamefont {N.~A.}\ \bibnamefont {Mortensen}},\
  }\href {\doibase https://doi.org/10.1002/adom.202101305} {\bibfield
  {journal} {\bibinfo  {journal} {Advanced Optical Materials}\ }\textbf
  {\bibinfo {volume} {9}},\ \bibinfo {pages} {2101305} (\bibinfo {year}
  {2021})}\BibitemShut {NoStop}%
\bibitem [{\citenamefont {Carmiggelt}\ \emph {et~al.}(2020)\citenamefont
  {Carmiggelt}, \citenamefont {Borst},\ and\ \citenamefont {van~der
  Sar}}]{Carmiggelt2020}%
  \BibitemOpen
  \bibfield  {author} {\bibinfo {author} {\bibfnamefont {J.~J.}\ \bibnamefont
  {Carmiggelt}}, \bibinfo {author} {\bibfnamefont {M.}~\bibnamefont {Borst}}, \
  and\ \bibinfo {author} {\bibfnamefont {T.}~\bibnamefont {van~der Sar}},\
  }\href {\doibase 10.1038/s41598-020-74376-3} {\bibfield  {journal} {\bibinfo
  {journal} {Scientific Reports}\ }\textbf {\bibinfo {volume} {10}},\ \bibinfo
  {pages} {17389} (\bibinfo {year} {2020})}\BibitemShut {NoStop}%
\bibitem [{\citenamefont {Oliver}\ \emph {et~al.}(2020)\citenamefont {Oliver},
  \citenamefont {Young}, \citenamefont {Krylyuk}, \citenamefont {Reinecke},
  \citenamefont {Davydov},\ and\ \citenamefont {Vora}}]{Oliver2020}%
  \BibitemOpen
  \bibfield  {author} {\bibinfo {author} {\bibfnamefont {S.~M.}\ \bibnamefont
  {Oliver}}, \bibinfo {author} {\bibfnamefont {J.}~\bibnamefont {Young}},
  \bibinfo {author} {\bibfnamefont {S.}~\bibnamefont {Krylyuk}}, \bibinfo
  {author} {\bibfnamefont {T.~L.}\ \bibnamefont {Reinecke}}, \bibinfo {author}
  {\bibfnamefont {A.~V.}\ \bibnamefont {Davydov}}, \ and\ \bibinfo {author}
  {\bibfnamefont {P.~M.}\ \bibnamefont {Vora}},\ }\href {\doibase
  10.1038/s42005-019-0277-7} {\bibfield  {journal} {\bibinfo  {journal}
  {Communications Physics}\ }\textbf {\bibinfo {volume} {3}},\ \bibinfo {pages}
  {10} (\bibinfo {year} {2020})}\BibitemShut {NoStop}%
\bibitem [{\citenamefont {Hao}\ \emph {et~al.}(2017)\citenamefont {Hao},
  \citenamefont {Xu}, \citenamefont {Wu}, \citenamefont {Nagler}, \citenamefont
  {Tran}, \citenamefont {Ma}, \citenamefont {Sch\"{u}ller}, \citenamefont
  {Korn}, \citenamefont {MacDonald}, \citenamefont {Moody},\ and\ \citenamefont
  {Li}}]{Hao2017}%
  \BibitemOpen
  \bibfield  {author} {\bibinfo {author} {\bibfnamefont {K.}~\bibnamefont
  {Hao}}, \bibinfo {author} {\bibfnamefont {L.}~\bibnamefont {Xu}}, \bibinfo
  {author} {\bibfnamefont {F.}~\bibnamefont {Wu}}, \bibinfo {author}
  {\bibfnamefont {P.}~\bibnamefont {Nagler}}, \bibinfo {author} {\bibfnamefont
  {K.}~\bibnamefont {Tran}}, \bibinfo {author} {\bibfnamefont {X.}~\bibnamefont
  {Ma}}, \bibinfo {author} {\bibfnamefont {C.}~\bibnamefont {Sch\"{u}ller}},
  \bibinfo {author} {\bibfnamefont {T.}~\bibnamefont {Korn}}, \bibinfo {author}
  {\bibfnamefont {A.~H.}\ \bibnamefont {MacDonald}}, \bibinfo {author}
  {\bibfnamefont {G.}~\bibnamefont {Moody}}, \ and\ \bibinfo {author}
  {\bibfnamefont {X.}~\bibnamefont {Li}},\ }\href {\doibase
  10.1088/2053-1583/aa70f9} {\bibfield  {journal} {\bibinfo  {journal} {2D
  Materials}\ }\textbf {\bibinfo {volume} {4}},\ \bibinfo {pages} {025105}
  (\bibinfo {year} {2017})}\BibitemShut {NoStop}%
\bibitem [{\citenamefont {Wang}\ \emph {et~al.}(2018)\citenamefont {Wang},
  \citenamefont {Chernikov}, \citenamefont {Glazov}, \citenamefont {Heinz},
  \citenamefont {Marie}, \citenamefont {Amand},\ and\ \citenamefont
  {Urbaszek}}]{Wangg2018}%
  \BibitemOpen
  \bibfield  {author} {\bibinfo {author} {\bibfnamefont {G.}~\bibnamefont
  {Wang}}, \bibinfo {author} {\bibfnamefont {A.}~\bibnamefont {Chernikov}},
  \bibinfo {author} {\bibfnamefont {M.~M.}\ \bibnamefont {Glazov}}, \bibinfo
  {author} {\bibfnamefont {T.~F.}\ \bibnamefont {Heinz}}, \bibinfo {author}
  {\bibfnamefont {X.}~\bibnamefont {Marie}}, \bibinfo {author} {\bibfnamefont
  {T.}~\bibnamefont {Amand}}, \ and\ \bibinfo {author} {\bibfnamefont
  {B.}~\bibnamefont {Urbaszek}},\ }\href {\doibase
  10.1103/RevModPhys.90.021001} {\bibfield  {journal} {\bibinfo  {journal}
  {Reviews of Modern Physics}\ }\textbf {\bibinfo {volume} {90}},\ \bibinfo
  {pages} {021001} (\bibinfo {year} {2018})}\BibitemShut {NoStop}%
\end{thebibliography}%

\end{document}